\documentclass[aps,prl,twocolumn,groupedaddress]{revtex4-2}
\usepackage{amsmath,amsthm,amssymb,mathtools,tikz,booktabs,tensor,slashed,dutchcal,upgreek}
\usepackage[mathscr]{eucal}
\usetikzlibrary{cd}

\newcommand{\mc}{\mathcal} \newcommand{\mf}{\mathfrak} \newcommand{\mb}{\mathbb} \newcommand{\on}{\operatorname}  \newcommand{\ms}{\mathscr}
\newcommand{\slot}{\;\cdot\;} \newcommand{\la}{\langle} \newcommand{\ra}{\rangle}
\newcommand{\gm}{{\mc G}} \newcommand{\di}{\slashed D} \newcommand{\tr}{\on{Tr}}

\DeclareMathOperator{\gr}{\mathcal{R}}

\allowdisplaybreaks

\begin{document}

%Title of paper
\title{Batalin--Vilkovisky formulation of the $\ms N=1$ supergravity in ten dimensions}

% repeat the \author .. \affiliation  etc. as needed
% \email, \thanks, \homepage, \altaffiliation all apply to the current
% author. Explanatory text should go in the []'s, actual e-mail
% address or url should go in the {}'s for \email and \homepage.
% Please use the appropriate macro foreach each type of information

% \affiliation command applies to all authors since the last
% \affiliation command. The \affiliation command should follow the
% other information
% \affiliation can be followed by \email, \homepage, \thanks as well.
%\author{}
%\email[]{Your e-mail address}
%\homepage[]{Your web page}
%\thanks{}
%\altaffiliation{}
%\affiliation{}

\author{Julian Kupka}
\email{j.kupka@herts.ac.uk}
\author{Charles Strickland-Constable}
\email{c.strickland-constable@herts.ac.uk}
\author{Fridrich Valach}
\email{f.valach@herts.ac.uk}
\affiliation{Department of Physics, Astronomy and Mathematics,
University of Hertfordshire, College Lane, Hatfield, AL10 9AB, United Kingdom}

%Collaboration name if desired (requires use of superscriptaddress
%option in \documentclass). \noaffiliation is required (may also be
%used with the \author command).
%\collaboration can be followed by \email, \homepage, \thanks as well.
%\collaboration{}
%\noaffiliation

\date{\today}

\begin{abstract}
We present a full Batalin--Vilkovisky action in the component field formalism for $\ms N=1$ supergravity in ten dimensions coupled to Yang--Mills multiplets.
\end{abstract}

%\maketitle must follow title, authors, abstract, and keywords
\maketitle
\section{Introduction}
  Ten dimensional $\ms N=1$ supergravity coupled to super Yang--Mills was introduced in the early 80's in \cite{Bergshoeff:1981um,Chapline:1982ww,Dine:1985rz}. Beyond serving as a ``mother'' theory for a host of lower-dimensional supergravity theories, it is particularly important for the low-energy description of the type I and heterotic superstring theory. Recently \cite{pre,short} the framework of generalised geometry was used as a convenient packaging tool to provide a simplified description of this theory (building on earlier works \cite{Siegel:1993th,CSW1,Garcia-Fernandez:2013gja,Coimbra:2014qaa}), in particular simplifying significantly the structure of the four-fermion terms in the action. %In the special case where the generalised metric (defined below) is frozen to be the identity operator one obtains a topological ``dilatonic supergravity'' theory, whose BV formulation was discovered in \cite{}.
  
  In this letter we make the next step forward and present a full Batalin--Vilkovisky (BV) action of the theory. We partially leverage the fact that the BV formulation was already constructed for a very special sector in the generalised-geometric moduli space, namely when the generalised metric (defined below) is frozen to be the identity operator. This leads to a topological theory (``dilatonic supergravity''), whose BV description was built in \cite{Kupka:2024tic}.
  
  Nevertheless, generalising this description to the nontopological case of physical supergravity is tricky. One complication is caused by the fact that the field space is naturally the total space of a vector bundle, and the fermionic fields (the dilatino, gravitino, and gaugino) do not correspond to coordinates on this field space, but rather describe elements of the fibers (this is a consequence of the fact that they are sections of the spinor bundle whose very definition requires a choice of metric). This problem is typically dealt with by passing to the vielbein description of the metric. Here we take a more direct geometric approach and simply work with the vector bundle structure directly. This leads to some significant simplifications --- for instance the algebra of local supersymmetries \eqref{nasty} does not feature any Lorentz terms on the RHS as these are cancelled by the term coming from the nonzero curvature on this vector bundle.
  
  Consequently the BV action \eqref{bv} is much simpler than one would apriori expect. Still, checking explicitly that it satisfies the classical master equation is not easy. Consequently we do not give a full proof of this fact but only provide various evidence in favour of its validity. A complete proof is left for a future work.
  
  Although immensely important for the purpose of quantisation, the BV analysis of supergravity has so far been mostly restricted to the $D=4$ case \cite{Baulieu:1990uv} (more recently see also \cite{Cattaneo:2024sfd,CFR}). To the best knowledge of the authors the present work is the first instance when the BV action for a higher-dimensional supergravity has been constructed in the background independent component field formalism (as opposed to the pure spinor superfield approach, see \cite{Cederwall:2022fwu} and references therein).
  
  We conclude the Introduction by highlighting the fact that the BV formulation of the $\ms N=1$ $D=10$ case is particularly interesting in that it is directly linked to the work \cite{Costello:2019jsy} of Costello--Li, since it provides the starting point for their twist of supergravity; for more details see the last section.

  %Some important preceding work in this direction includes the BV description of the $D=4$ supergravity
  
%  This can be seen as a marriage of the generalised-geometric formulation of \cite{pre} and of the BV analysis of the topological ``dilatonic supergravity'' \cite{}, where the latter corresponds to a particular ``limit'' of the supergravity theory when the generalised metric (discussed below) is frozen to be the identity operator.
  
\section{Bosonic field content}
  We first recall the generalised-geometric description of supergravity, based on \cite{CSW1,pre}. We refer the reader to these works for more details and conventions.
  
  The theory itself is defined in terms of a transitive Courant algebroid $E$ \cite{liu1997manin,let} over a ten-dimensional base space $M$. Locally, this is given by a vector bundle
  \begin{equation}\label{loc}
    E\cong_{\text{loc}} TM\oplus T^*M\oplus (\mf g\times M),
  \end{equation}
  with $\mf g$ a Lie algebra with an invariant pairing denoted by $\tr$. Sections of $E$ thus correspond to formal sums of a vector field, a 1-form, and a $\mf g$-valued function. This structure is equipped with a bracket, pairing, and a map $a\colon E\to TM$ given by
  \begin{equation*}
    \begin{aligned}
      [X\!+\alpha+s,Y\!+\beta+t]&=L_XY+(L_X\beta-i_Yd\alpha+\tr t\,ds)\\
      &\quad+(L_Xt-L_Ys+[s,t]_\mf g),\\
      \langle X\!+\alpha+s,Y\!+\beta+t\ra&=\alpha(Y)+\beta(X)+\tr st\\
      a(X+\alpha+s)&=X.
    \end{aligned}
  \end{equation*}
%  A particularly useful operator on a Courant algebroid is the generalised Lie derivative $\ms L_u$, for $u\in\Gamma(E)$. This acts on other sections by $\ms L_uv=[u,v]$, on functions on $M$ by $\ms L_uf=a(u)f$, and also naturally extends on 
  
  Let $H$ denote the line bundle of half-densities on $M$, and $\ms H^*$ be the space of its invertible (i.e.\ everywhere nonvanishing) sections.
  The bosonic field content of the theory then consists of the following fields:
  \begin{itemize}
    \item a generalised metric $\gm$, i.e.\ a symmetric endomorphism $E\to E$ satisfying $\gm^2=1$,
    \item an invertible half-density $\sigma\in \ms H^*$.
  \end{itemize}
  The generalised metric induces an orthogonal splitting $E=C_+\oplus C_-$ into its $\pm1$ eigenbundles. We will denote the frames of $C_+$ and $C_-$ by $e_a, e_b,\dots$ and $e_\alpha, e_\beta,\dots$, respectively. We shall make a further assumption that
  \begin{itemize}
    \item $\la \slot,\slot\ra|_{C_+}$ has signature $(9,1)$ and admits spinors
    \item $a|_{C_+}\colon C_+\to TM$ is an isomorphism.
  \end{itemize}
  Denote the space of such generalised metrics by $\ms M$. The last two conditions provide the bridge to the ordinary description of the field content, as under the identification \eqref{loc} any $C_+$ takes the form
  \begin{equation}
    \{x+(i_xg+i_xB-\tfrac12\tr A\,i_xA)+i_xA\mid x\in TM\}\subset E
  \end{equation}
  for some Lorentzian metric $g$, Kalb--Ramond 2-form $B$, and $G$-connection 1-form $A$. The dilaton function $\phi$ is encoded in $\sigma$ via
  \begin{equation}
    \sigma^2=\sqrt{|g|}e^{-2\phi},
  \end{equation}
  where $\sqrt{|g|}$ stands for the standard metric density.
  
    Consider now the ``tautological'' bundle $\ms C_+\to \ms M$, whose fiber at $\gm$ is the space $\Gamma(C_+)$. Any small change $\gm \leadsto \gm':=\gm+\delta\gm$ induces a small deformation of the subbundle $C_+\leadsto C_+'$. Since the orthogonal projection $E\to C_+$ gives an isomorphism $C_+'\to C_+$ (see the following picture), we have an identification of the nearby fibers of $\ms C_+$, i.e.\ a connection.
  
  
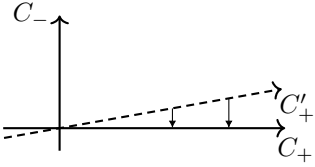
\begin{center}
  \begin{tikzpicture}[scale=1.5]
    \draw[->, thick] (-.5,0) -- (2,0) node[anchor=north,xshift=4pt] {$C_+$};
    \draw[->, thick] (0,-.2) -- (0,1) node[anchor=east] {$C_-$};
    \draw[densely dashed, rotate=10, ->, thick] (-.5,0) -- (2,0) node[anchor=north,xshift=6pt,yshift=2pt] {$C_+'$};
    \draw[-{Latex[length=1mm,width=1mm]}] (1,.175) -- (1,0);
    \draw[-{Latex[length=1mm,width=1mm]}] (1.5,.2625) -- (1.5,0);
  \end{tikzpicture}
  \end{center}
  
  A straightforward calculation shows that the curvature of this connection is
  \begin{equation}\label{curv}
    F(\delta_1\gm,\delta_2\gm)=\tfrac14 [\delta_1\gm,\delta_2\gm]\colon \Gamma(C_+)\to \Gamma(C_+),
  \end{equation}
  where $\delta_{1,2}\gm$ are two infinitesimal variations of $\gm$, i.e.\ vectors at $T_\gm\ms M$. Analogously we obtain a connection and curvature on $\ms C_-\to \ms M$.

%Furthermore, both $\ms C_\pm$ carry a non-degenerate inner product given by
%  \begin{equation}
%    \int_M\la u_\pm,v_\pm\ra\sigma^2
%  \end{equation}
%  for $u_\pm,v_\pm\in \Gamma(C_\pm)$, and this inner product is preserved by the connection.
%also on $T\ms M\to \ms M$, whose fibers over $\gm\in\ms M$ can be naturally identified with $\Gamma(C_+^*\otimes C_-\otimes H^2)$. It can be shown that the latter is in fact the Levi-Civita connection for the canonical metric on $\ms M$, but we will not use this fact explicitly in this letter.
  
\section{Fermions and supersymmetry}
  Denoting the Majorana--Weyl spinor bundles for $C_+$ by $S_\pm$, the fermionic field content of the theory is
  \begin{equation}
    \rho\in\Gamma(\Pi S_+\otimes H),\qquad \psi\in\Gamma(\Pi S_-\otimes C_-\otimes H),
  \end{equation}
  where $\Pi$ denotes the parity shift. As shown in \cite{pre}, these fields encode the usual dilatino and gravitino+gaugino, respectively. Note that we define the fermions as half-densities.
  
  Since $\rho$ and $\psi$ are sections of bundles which themselves depend on the generalised metric, the classical field space has the structure of (the total space of) a vector bundle
  \begin{equation}
    \ms S_0 \to \ms M\times \ms H^*,
  \end{equation}
  whose fibre at $(\gm,\sigma)$ is
  \begin{equation}
    \Gamma(\Pi S_+\otimes H)\times\Gamma(\Pi S_-\otimes C_-\otimes H).
  \end{equation}
  Since the bundles $S_\pm$ are naturally associated to $C_+$, it follows that $\ms S_0$ carries a connection inherited from the ones on $\ms C_\pm \to \ms M$. Its curvature is
  \begin{equation}\label{curvspin}
    \begin{aligned}
      F(\delta_1\gm,\delta_2\gm)\rho&=\tfrac18\delta_1\gm_a{}^\beta\delta_2\gm_{b\beta}\gamma^{ab}\rho,\\
      F(\delta_1\gm,\delta_2\gm)\psi^\alpha&=\tfrac18\delta_1\gm_a{}^\beta\delta_2\gm_{b\beta}\gamma^{ab}\psi^\alpha\\
      &\qquad \qquad + \tfrac12\delta_{[1}\gm_a{}^\alpha\delta_{2]}\gm^a{}_\beta \psi^\beta.
    \end{aligned}
  \end{equation}
  Note that this has the form of a Lorentz transformation.
  
  In order to write down kinetic terms for the fermions we again recall the construction from \cite{CSW1,pre}. A generalised connection $D$ \footnote{A generalised connection is a map $D\colon \Gamma(E)\times\Gamma(E)\to\Gamma(E)$ satisfying $D_{fu}v=fD_uv$, $D_u(fv)=fD_uv+(a(u)f)v$, and $D\la \slot,\slot\ra=0$. Any such connection also naturally acts on half-densities via $D_u\sigma=L_{a(u)}\sigma-\tfrac12 \sigma \on{tr}(Du)$, where $L$ is the usual Lie derivative.} is said to belong to the class $LC(\gm,\sigma)$ if it is torsion-free and preserves both $\gm$ and $\sigma$. Such connection exist but are not unique \cite{Garcia-Fernandez:2016ofz}; however, there exist objects constructed out of $\gm$, $\sigma$, and $D\in LC(\gm,\sigma)$, which are independent of the choice of the representative $D\in LC(\gm,\sigma)$ and thus only depend on $\gm$ and $\sigma$. The most important are
  \begin{itemize}
    \item the generalised scalar curvature $\gr$
    \item the generalised Ricci tensor $\gr_{a\beta}$
    \item the Dirac operator $\di=\gamma^aD_a$ in $\di\rho$ and $\di\psi^\alpha$
    \item the operator $D_\alpha$ in $D_\alpha\rho$ and $D_\alpha\psi^\alpha$
    \item the operator $D$ when acting on any $f\in C^\infty(M)$.
  \end{itemize}
  
  This allows us to construct the following action functional $S_0$ \cite{pre} on the space $\ms S_0$:
  \begin{equation}\label{a}
      \begin{aligned}
        S_0&=\smash{\int_M}\mc R\sigma^2+\bar\psi_{\alpha}\slashed D\psi^{\alpha}+\bar\rho\slashed D\rho+2\bar\rho D_{\alpha}\psi^{\alpha}\\
        &\qquad\qquad -\tfrac1{768}\sigma^{-2}(\bar\psi_{\alpha}\gamma_{abc}\psi^{\alpha})(\bar\rho\gamma^{abc}\rho)\\
        &\qquad\qquad-\tfrac1{384}\sigma^{-2}(\bar\psi_{\alpha}\gamma_{abc}\psi^{\alpha})(\bar\psi_{\beta}\gamma^{abc}\psi^{\beta}),
      \end{aligned}
    \end{equation}
  which is invariant under the local supersymmetries, i.e.\
  \begin{equation}\label{s}
    \begin{aligned}
        \delta_\epsilon \mc G_{ab}&=\delta_\epsilon\mc G_{\alpha\beta}=0,\;\; \delta_\epsilon\mc G_{a\beta}=\delta_\epsilon\mc G_{\beta a}=\tfrac12\sigma^{-2}\bar \epsilon\gamma_a\psi_{\beta}\\
        \delta_\epsilon\sigma&=\tfrac18\sigma^{-1}(\bar\rho\epsilon)\\
        \delta_\epsilon\rho&=\slashed D\epsilon+\tfrac1{192}\sigma^{-2}(\bar\psi_{\alpha}\gamma_{abc}\psi^{\alpha})\gamma^{abc}\epsilon\\
        \delta_\epsilon\psi_{\alpha}&=D_{\alpha}\epsilon+\tfrac18\sigma^{-2}(\bar\psi_{\alpha}\rho)\epsilon+\tfrac18\sigma^{-2}(\bar\psi_{\alpha}\gamma_a\epsilon)\gamma^a\rho,
    \end{aligned}
    \end{equation}
    The supersymmetry parameter $\epsilon$ is here a function on $\ms S_0$, which for any given field configuration $(\gm,\sigma,\rho,\psi)$ takes value in $\Gamma(\Pi S_-\otimes H)$ (note that this  bundle itself depends on $\gm$). Formulas \eqref{s} thus define a vector field $\delta_\epsilon$ on $\ms S_0$ --- more precisely the first and second pair of formulas express the horizontal and vertical parts of this vector field, respectively. This is the general meaning of the formulas for supersymmetry variations.\footnote{In other words, notice that $\delta_\epsilon\rho$ corresponds to the difference of two sections of two different bundles (corresponding to $\gm$ and $\gm+ \delta_\epsilon\gm$, respectively). In order to make sense of this one needs to provide the identification of these two bundles, which is where the connection on $\ms S_0$ is used.}
    
  Similarly, we note that any section $\zeta\in\Gamma(E)$ induces an infinitesimal automorphism of $E$. This is usually expressed via the generalised Lie derivative operator $\ms L_\zeta$ and again produces a vector field $\delta_\zeta$ on $\ms S_0$ which preserves the functional $S_0$. Note that this action is reducible, as $\ms L_{Df}=0$ for any $f\in C^\infty(M)$.
  
  Let us now turn to the algebra of the symmetries.
  As usual, its most interesting part corresponds to the commutator of two supersymmetries. On the bosonic fields we have
  \begin{equation}
    [\delta_{\epsilon_1},\delta_{\epsilon_2}]=\delta_\epsilon+\delta_\zeta,
  \end{equation}
  where we defined
  \begin{equation}
    \zeta^a:=\tfrac14\sigma^{-2}\bar\epsilon_2\gamma^a\epsilon_1,\quad \epsilon:=-\tfrac12\slashed\zeta \rho+\delta_{\epsilon_1}\epsilon_2-\delta_{\epsilon_2}\epsilon_1,
  \end{equation}
  while on the fermions
  \begin{equation}\label{nasty}
    \begin{aligned}
      [\delta_{\epsilon_1},\delta_{\epsilon_2}]\rho&=\delta_{\epsilon}\rho+\delta_\zeta\rho-\tfrac12\slashed \zeta(\di \rho+\dots),\\
      [\delta_{\epsilon_1},\delta_{\epsilon_2}]\psi^\alpha&=\delta_\epsilon\psi^\alpha+\delta_\zeta\psi^\alpha+(\tfrac14\epsilon_{[2}\bar\epsilon_{1]}-\tfrac12\slashed \zeta)(\di \psi^\alpha+\dots),
    \end{aligned}
  \end{equation}
  where the last parentheses contain the equations of motion of $\rho$ and $\psi$, respectively. Note that this provides a significant simplification when compared to the usual formulae (cf.\ \cite{Bergshoeff:1981um}).
  
  As usual, the calculations leading to \eqref{nasty} are relatively lengthy and involve a generous handful of Fierz identities. Notably, since the formulas \eqref{s} really correspond to horizontal and vertical parts of the vector field $\delta_\epsilon$ on $\ms S_0$, the corresponding commutator on fermions picks up an extra term (in addition to the ``naive'' commutator of variations), coming from the curvature \eqref{curvspin}. This removes all the terms which look like Lorentz transformations and which would be present in the vielbein formulation. (Note that Lorentz transformations are not expected to appear in \eqref{nasty} as they are \emph{not} symmetries of the \emph{metric tensor formulation} of the theory.) We postpone the details of the calculation to a future work \cite{commutator}.
  
\section{BV field space}
  To construct the BV space we start with the classical field space, add ghosts and ghosts for ghosts corresponding to local symmetries, and then adjoin the corresponding antifields. This yields the BV space
  \begin{equation}
    \ms F_{BV}:=T^*[-1]\ms S,
  \end{equation}
  where $\ms S\to \ms M\times\ms H^*$ is the vector bundle whose fiber at $(\gm,\sigma)$ is
  \begin{equation}\label{fiber}
    \begin{aligned}
      \Gamma(\Pi S_+&\otimes H)\times\Gamma(\Pi S_-\otimes C_+\otimes H)\\
      &\times \Gamma(\Pi S_-\otimes H)[1]\times \Gamma(E)[1]\times C^\infty(M)[2].
    \end{aligned}
  \end{equation}
  Here $[n]$ signifies the degree shift and corresponds to the ghost number. The overall parity is the sum of the superdegree (bosonic/fermionic) and the parity of the ghost degree.
  Elements of the fiber \eqref{fiber} correspond to the fermionic fields $\rho$ and $\psi$, the supersymmetry ghost $e$, the diffeomorphism ghost $\xi$, and the ghost for ghost $f$, respectively. To summarise, our field content up to now consists of
  \begin{equation}
    \begin{aligned}
      \gm&\in \Gamma(E^*\otimes E)\text{ s.t.\ }\gm^2=1\text{ and }\gm^*=\gm\\
      \sigma&\in \Gamma(H)\text{ everywhere nonvanishing}\\
      \rho&\in\Gamma(\Pi S_+\otimes H)\\
      \psi&\in\Gamma(\Pi S_-\otimes C_+\otimes H)\\
      e&\in\Gamma(\Pi S_-\otimes H)[1]\\
      \xi&\in\Gamma(E)[1]\\
      f&\in C^\infty(M)[2]
    \end{aligned}
  \end{equation}
  In particular the fields $\gm$, $\sigma$, $e$, and $f$ are even and the rest is odd.
  
  The space $\ms F_{BV}$ also includes the dual antifields, whose most convenient description is as follows. We start by noting that, following the discussion above, $\ms S$ carries a natural connection and hence a splitting of its tangent spaces into horizontal and vertical parts. This gives an identification
  \begin{equation}
    T^*[-1]\ms S\cong \pi^* (T^*[-1](\ms M\times\ms H^*))\oplus \pi^*\ms S^*[-1],
  \end{equation}
  of bundles over $\ms S$, where $\pi\colon \ms S\to \ms M\times\ms H^*$ is the projection. We will describe the fibers of the first and second summand by the dual coordinates $\gm^*$, $\sigma^*$, and $\psi^*$, $\rho^*$, $\xi^*$, $e^*$, $f^*$, respectively. More concretely, for any configuration $(\gm,\sigma,\psi,\rho,\xi,e,f)$ we have
  \begin{equation}
    \begin{aligned}
      \gm^*&\in T^*_\gm[-1]\ms M\cong \Gamma(C_+\otimes C_-\otimes H^2)[-1]\\
      \sigma^*&\in\Gamma(H)[-1]\\
      \psi^*&\in\Gamma(\Pi S_+\otimes C_-\otimes H)[-1]\\
      \rho^*&\in\Gamma(\Pi S_-\otimes H)[-1]\\
      \xi^*&\in\Gamma(E\otimes H^2)[-2]\\
      e^*&\in\Gamma(\Pi S_+\otimes H)[-2]\\
      f^*&\in\Gamma(H^2)[-3]
    \end{aligned}
  \end{equation}
  Here we used the fact that infinitesimal deformations of $\gm$ correspond to deformations of $C_+$; and any nearby deformed $C'_+$ is the graph of a vector bundle map $C_+\to C_-$ (see the picture above). We also used the identifications $C_\pm\cong C_\pm^*$, $E^*\cong E$, and $S_\pm^*\cong S_\mp$.
  Note that $\psi^*$, $\rho^*$, and $\xi^*$ are even and the rest is odd.
  
\section{BV action}
  We now claim that the BV extension of the supergravity action \eqref{a} is
    \begin{align}
    S&=\smash{\int_M}\mc R\sigma^2+\bar\psi_{\alpha}\slashed D\psi^{\alpha}+\bar\rho\slashed D\rho+2\bar\rho D_{\alpha}\psi^{\alpha}\nonumber\\
        &\qquad\quad -\tfrac1{768}\sigma^{-2}(\bar\psi_{\alpha}\gamma_{abc}\psi^{\alpha})(\bar\rho\gamma^{abc}\rho)\nonumber\\
        &\qquad\quad-\tfrac1{384}\sigma^{-2}(\bar\psi_{\alpha}\gamma_{abc}\psi^{\alpha})(\bar\psi_{\beta}\gamma^{abc}\psi^{\beta})\nonumber\\
        &\qquad\quad+\sigma^*[\ms L_\xi \sigma-\tfrac18 \sigma^{-1}(\bar\rho e)]\nonumber\\
        &\qquad\quad+\gm^*_{a\beta}[(\ms L_\xi\gm)^{a\beta}+\tfrac12\sigma^{-2}(\bar e\gamma^a\psi^{\beta})]\nonumber\\
        &\qquad\quad+\bar\rho^*[\ms L_\xi \rho+\di e+\tfrac1{192}\sigma^{-2}(\bar\psi_\beta\gamma_{abc}\psi^\beta)\gamma^{abc}e]\nonumber\\
        &\qquad\quad+\bar\psi^*_{\beta}[(\ms L_\xi \psi)^{\beta}+D^{\beta}e+\tfrac18\sigma^{-2}(\bar\psi^{\beta}\rho)e\label{bv}\\
        &\qquad\qquad\qquad-\tfrac18\sigma^{-2}(\bar\psi^{\beta}\gamma_ae)\gamma^a\rho]\nonumber\\
        &\qquad\quad+\bar e^*[\ms L_\xi e+\tfrac1{16}\sigma^{-2}(\bar e\gamma_a e)\gamma^a\rho]\nonumber\\
        &\qquad\quad+\la \xi^*,\ms D\!f+\tfrac12\ms L_\xi\xi\ra-\tfrac18\xi^{*}_a\sigma^{-2}(\bar e\gamma^a e)\nonumber\\
        &\qquad\quad+\tfrac12f^*(\ms L_\xi f+\tfrac18\sigma^{-2}(\bar e\gamma_a e)\xi^a-\tfrac16\la \xi,\ms L_\xi\xi\ra)\nonumber\\
        &\qquad\quad-\tfrac1{64}\sigma^{-2}(\bar e\gamma_a e)(\bar\psi^*_\beta\gamma^a\psi^{*\beta})-\tfrac1{32}\sigma^{-2}(\bar e\psi^*_{\beta})(\bar e\psi^{*\beta})\nonumber\\
        &\qquad\quad -\tfrac1{64}\sigma^{-2}(\bar e\gamma_a e)(\bar\rho^*\gamma^a\rho^*).\nonumber
    \end{align}
    
    Following the usual BV machinery the form of this action is essentially read off from what was discussed before: the linear terms in antifields include both \eqref{s} and the generalised diffeomorphisms, as well as the ``structure coefficients'' of the symmetry algebra (first part of the RHS of \eqref{nasty}), while the terms quadratic in antifields quantify the failure of the symmetries to close off-shell (last part of the RHS of \eqref{nasty}). Finally, an important nontrivial check is provided by the fact that when taking $\gm=1$ (and after a constant rescaling of $\rho$) the expression \eqref{bv} matches the BV action for the dilatonic supergravity \cite{Kupka:2024tic}. This can in particular be used to determine the terms in \eqref{bv} containing the ghost-for-ghost $f$, which account for the reducibility of our description of generalised diffeomorphisms (which is the same regardless of whether $\gm=1$ or not).\footnote{As seen in \cite{pre} this corresponds reducibility of the gauge transformations of the $B$-field.} Our result also structurally matches the $D=4$ supergravity BV analysis in \cite{Baulieu:1990uv}.
    
    That being said, we note that although highly suggestive, the above arguments do not provide a full proof that the classical master equation is indeed satisfied. However, even after performing additional nontrivial checks (which are too lengthy to report on here) we have not found any indication that the formula \eqref{bv} is incomplete or incorrect in any way and hence we are highly confident in its validity. The full proof of this fact is left for a future work.

\section{Conclusions and outlook}
  %Having determined the BV formulation of the $\ms N=1$ 10D supergravity we can now proceed to various applications and outlook.
  %
  %Let us first consider the present work in the context of the twisting of supergravity.
  We have found the BV action for the $\ms N=1$ supergravity in 10 dimensions, in general coupled to a super Yang--Mills sector. It looks somewhat likely that with further effort one might succeed in performing a similar BV analysis for the type II supergravity. One should also be able to derive the corresponding results for the lower-dimensional supergravities via consistent truncations.
  Using our results one could proceed to look for a perturbative solution to the quantum master equation in order to investigate the quantum nature of the supergravity theories.
  It would also be interesting to relate the present BV formulation to superstring field theory. We leave these questions for future work.
  
  In \cite{Costello:2016mgj} Costello and Li suggested a procedure of twisting supergravity. Their twist starts by taking the BV formulation of supergravity and then expanding the BV action $S$ around its critical point \footnote{By a point we here mean a point in the even part $\ms F_{BV}^{\text{even}}$ of the supermanifold $\ms F_{BV}$, i.e.\ a configuration with all odd fields vanishing.} which has a nonzero value of the supersymmetry ghost $e$. In particular it was conjectured in \cite{Costello:2019jsy} that the (holomorphic) twist of type I supergravity on a Calabi--Yau 5-fold is described by the $\mb Z_2$-fixed locus of the BCOV theory \cite{Bershadsky:1993cx} coupled to the $SO(32)$ holomorphic Chern--Simons theory.
   The present work could be used to put this conjecture on more solid ground by completing the BV description of (the two-derivative part of) its starting point, i.e.\ type I supergravity.
   % () --- the classical BV formulation of the $\ms N=1$ theory in the component field formulation.
  
  Following this philosophy, let us look more closely at the critical points of the BV action \eqref{bv} (for any gauge group). The corresponding equations are easy to find and are shown in \eqref{crit}. In particular setting to zero all the fields except for $\gm$, $\sigma$, and $e$ these equations reduce to
  \begin{equation}\label{cy}
    \gr_{a\beta}=\di e= D_\alpha e= \bar e\gamma^a e=0.
  \end{equation}
  This is the condition for the background $(\gm,\sigma)$ to be supersymmetric, with the extra requirement that $\bar e\gamma^a e=\gr_{a\beta}=0$. (Note that the vanishing of the generalised scalar curvature $\gr$ follows from the Lichnerowitz formula \cite{CSW1,pre}.) The equations \eqref{crit} can be therefore regarded as a generalisation thereof. This in particular suggest an interesting modification of the condition for the existence of a parallel spinor (leading in the classical case to e.g.\ Calabi--Yau manifolds) to one of the form
  \begin{equation}
    D_\alpha e=\tfrac1{16}\sigma^{-2}e(\bar\psi^*_\alpha e).
  \end{equation}
  We recall here that both spinors $e$ and $\psi^*$ are even (commuting).

\appendix
\section{Appendix: Critical points of the BV action}
  We are interested in critical points of $S$ on $\ms F_{BV}^{\text{even}}$. Since all the terms in $S$ contain an even number of odd fields, we can first consistently set them all to zero before performing the variation. The equation of motion for $\xi^*$ then becomes
  \begin{equation}
    D_af-\tfrac18\sigma^{-2}\bar e\gamma_a e=0,\qquad  D_\alpha f=0.
  \end{equation}
  Since in our setup $a|_{C_-}$ is surjective it follows that
  \begin{equation}
    df=0,\qquad \bar e\gamma_a e=0.
  \end{equation}
  The equation of motion for $\rho^*$ then reduces to
  \begin{equation}
    \di e=0.
  \end{equation}
  A straighforward (and very short) calculation using formulas from \cite{pre} then shows that the remaining equations are
  \begin{align}
      0&=\gr_{a\alpha}\!\sigma^2+\tfrac14\bar\rho^*\gamma_aD_\alpha e-\tfrac14\bar e\gamma_aD_\alpha\rho^*-\tfrac12\bar\psi^*_\alpha D_a e\nonumber\\
      &\qquad-\tfrac14\bar\psi^*_\alpha\gamma_{ab}D^be-\tfrac14\bar e\gamma_{ab}D^b\psi^*_\alpha\nonumber\\
      0&=\gr\!\sigma^2+\tfrac12\bar\psi^*_\alpha D^\alpha e-\tfrac12 \bar eD^\alpha \psi^*_\alpha+\tfrac1{32}\sigma^{-2}(\bar e\psi^*_\alpha)(\bar e\psi^{*\alpha})\nonumber\\
      0&=D_\alpha e-\tfrac1{16}\sigma^{-2}e(\bar\psi^*_\alpha e)\label{crit}\\
      0&=D^\alpha \psi^*_\alpha-\di\rho^*-\tfrac14\sigma^{-2}\xi^*_a\gamma^ae-\tfrac1{32}\sigma^{-2}\gamma_a e(\bar\psi^*_\alpha\gamma^a\psi^{*\alpha})\nonumber\\
      &\qquad -\tfrac1{16}\sigma^{-2}\psi^*_\alpha(\bar e\psi^{*\alpha})-\tfrac1{32}\sigma^{-2}\gamma_a e(\bar\rho^*\gamma^a\rho^*)\nonumber\\
      0&=D^a\xi^*_a+D^\alpha \xi^*_\alpha.\nonumber
  \end{align}

%   \begin{equation}\label{crit}
%     \begin{aligned}
%       0&=\gr_{a\alpha}\!\sigma^2+\tfrac14\bar\rho^*\gamma_aD_\alpha e-\tfrac14\bar e\gamma_aD_\alpha\rho^*-\tfrac12\bar\psi^*_\alpha D_a e\\
%       &\qquad-\tfrac14\bar\psi^*_\alpha\gamma_{ab}D^be-\tfrac14\bar e\gamma_{ab}D^b\psi^*_\alpha\\
%       0&=\gr\!\sigma^2+\tfrac12\bar\psi^*_\alpha D^\alpha e-\tfrac12 \bar eD^\alpha \psi^*_\alpha+\tfrac18\xi^*_aV^a\\
%       &\qquad+\tfrac1{64}\bar\psi^*_\alpha \slashed V\psi^{*\alpha}+\tfrac1{64}\bar\rho^*\slashed V\rho^*+\tfrac1{32}\sigma^{-2}(\bar e\psi^*_\alpha)(\bar e\psi^{*\alpha})\\
%       0&=\di e-\tfrac1{32}\slashed V\rho^*\\
%       0&=D_\alpha e-\tfrac1{32}\slashed V\psi^*_\alpha-\tfrac1{16}\sigma^{-2}e(\bar\psi^*_\alpha e)\\
%       0&=D^\alpha \psi^*_\alpha-\di\rho^*-\tfrac14\sigma^{-2}\xi^*_a\gamma^ae-\tfrac1{32}\sigma^{-2}\gamma_a e(\bar\psi^*_\alpha\gamma^a\psi^{*\alpha})\\
%       &\qquad -\tfrac1{16}\sigma^{-2}\psi^*_\alpha(\bar e\psi^{*\alpha})-\tfrac1{32}\sigma^{-2}\gamma_a e(\bar\rho^*\gamma^a\rho^*)\\
%       0&=D_af-\tfrac18V_a = D_\alpha f=D^a\xi^*_a+D^\alpha \xi^*_\alpha,
%     \end{aligned}
%   \end{equation}
  
% If you have acknowledgments, this puts in the proper section head.
\begin{acknowledgments}
C.S.-C.~and F.V.~are supported by an EPSRC New Investigator Award, grant number EP/X014959/1. No new data was collected or generated during the course of this research. 
\end{acknowledgments}

% Create the reference section using BibTeX:
\bibliography{citations}

\end{document}